\documentclass[preprint,aps,superscriptaddress,nofootinbib]{revtex4}
\pdfoutput=1
\usepackage{graphicx}
\usepackage{amsmath}
\usepackage[usenames]{color}
\usepackage{amssymb}
\usepackage[T1]{fontenc}

\usepackage[caption=false]{subfig}
\newcommand{\be}{\begin{equation}}
\newcommand{\ee}{\end{equation}}
\newcommand{\bea}{\begin{eqnarray}}
\newcommand{\eea}{\end{eqnarray}}
\newcommand{\pr}{\partial}
\newcommand{\nno}{\nonumber}
\newcommand{\bse}{\begin{subequations}}
\newcommand{\ese}{\end{subequations}}

\newcommand{\LL}{\mathcal{L}}

\def\be{\begin{equation}}
\def\ee{\end{equation}}
\usepackage[T1]{fontenc}
\usepackage[latin1]{inputenc}

\def\M{{\mathcal M}}

\def\be{\begin{equation}}
\def\ee{\end{equation}}

\def\a{\alpha}
\def\b{\beta}
\def\g{\gamma}
\def\d{\delta}
\def\e{\epsilon}

\def\r{\rho}
\def\l{\lambda}
\def\k{\kappa}\def\m{\mu}\def\n{\nu}\def\s{\sigma}\def\l{\lambda}

\def\bg{\bar{g}}
\def\beq{\begin{eqnarray}}\def\eeq{\end{eqnarray}}
\def\ba#1\ea{\begin{align}#1\end{align}}
\def\bg#1\eg{\begin{gather}#1\end{gather}}
\def\bm#1\em{\begin{multline}#1\end{multline}}
\def\bmd#1\emd{\begin{multlined}#1\end{multlined}}
\def\bea{\begin{eqnarray}}
\def\eea{\end{eqnarray}}

\def\a{\alpha}
\def\b{\beta}

\def\d{\delta}

\def\e{\epsilon}

\def\g{\gamma}

\def\k{\kappa}
\def\l{\lambda}

\def\m{\mu}
\def\n{\nu}

\def\r{\rho}
\def\s{\sigma}

\def\z{\zeta}
\def\del{\partial}

\def\fr{\frac}

\begin{document}

\title{Constraining scalar-Gauss-Bonnet Inflation by Reheating, Unitarity and PLANCK}

\author{Srijit Bhattacharjee$^{\, a, b}$}
\email{srijuster@gmail.com}
\affiliation{$^{a\,}$Institute of Physics\\
Sachivalaya Marg, Bhubaneshwar, Odisha, India 751005\\ 
\vspace{1mm}
$^{b \,}$Indian Institute of Technology Gandhinagar\\
Palaj, Gandhinagar, Gujarat, India 382355\\ }
\vspace{2mm}
\author{Debaprasad Maity}
\email{debu.imsc@gmail.com}
\affiliation{Indian Institute of Technology Guwahati\\
Guwahati, Assam, India 781039 \\}
\vspace{2mm}
\author{Rupak Mukherjee$^{\,c, d}$}
\email{rupakmukherjee01@gmail.com}
\affiliation{$^{c\,}$Institute for Plasma Research\\
Gandhinagar, Near Indira Bridge, Gujarat, India 382428\\$^{d\,}$Homi Bhabha National Institute, Training School Complex, Anushakti Nagar, Mumbai 400085, India}


\begin{abstract}
We revisit the inflationary dynamics in detail for theories with Gauss-Bonnet gravity coupled to scalar functions, in light of the Planck data. Considering the chaotic inflationary scenario, we constrain the parameters of two models involving inflaton-Gauss-Bonnet coupling by current Planck data. For non zero inflaton-Gauss-Bonnet coupling $\beta$, an inflationary analysis provides us a big cosmologically viable region in the space of $(m,\beta)$, where $m$ is the mass of inflaton. However, we study further on constraining $\beta$ arising from reheating considerations and unitarity of tree-level amplitude involving we have studied the constraints on $\beta$ arising from reheating considerations and unitarity of tree level amplitude involving $2$ graviton $\rightarrow 2$ graviton ($h h\rightarrow hh$) scattering. Our analysis, particularly on reheating significantly reduces the parameter space of $(m,\beta)$ for all models. he quadratic Gauss-Bonnet coupling parameter turns out to be more strongly constrained than that of the linear coupling. For the linear Gauss-Bonnet coupling function, we obtain  $\beta \lesssim 10^3$, with the condition $\beta (m/M_P)^2 \simeq 10^{-4}$. However, study of the Higgs inflation scenario in the presence of Gauss-Bonnet term turned out to be strongly disfavored.   

\end{abstract}

\maketitle

\newpage
\section{Introduction}\label{intro}

Inflation is assumed to have happened at the very early stage of the evolution of our universe. 
Increasingly precise cosmological observations in recent past have been providing us a clear hint 
towards this paradigm of theoretical cosmology \cite{guth,linde1,steinhardt}. Although a wide variety of cosmological models have been studied assuming the inflationary paradigm, any fundamental principle implying the mechanism itself as well as its driving source is still not well understood. Almost all the inflationary models are operative in an energy scale that is higher than at least the grand unified theory (GUT) scale, beyond which the nature of our physical laws is unclear. At energies higher than this scale, new degrees of freedom may turn on, and we have to consider a completely new theory that will describe the dynamics or nature of the physical laws. String theory is believed to be one such candidate that describes the physics up to a scale where the quantum effects of gravity are dominating. However, if we take a low-energy limit of such a theory, usually there exists a sufficient amount of symmetries that may be enough to predict the physical observables at this low-energy scale. This is the essence of the effective field theoretic point of view. In the effective field theoretic framework, one can ignore the degrees of freedom that may emerge at a higher energy scale and consider only the terms that are relevant at the energy scale under consideration, below which the theory works \cite{Donoghue}. This framework has been proved to be a very powerful tool to analyze the physics at energies closer to (but sufficiently below) the Planck scale (the scale at which quantum gravity effects set in). Many inflationary models available today take this route to analyze the dynamics of the Universe at the time of inflation. In this paper, we have considered a class of such theories, namely, the Gauss-Bonnet theory coupled with functions of a scalar field, and investigated if these theories can successfully predict the inflationary dynamics compatible with present observational constraints on the parameters of these theories. 

As already mentioned above, the prime motivation behind considering higher derivative gravity theories is its effective low-energy interpretation in the framework of string theory. The Gauss-Bonnet term is known to be generated as a low-energy effective action of heterotic string theory \cite{Zweibach}. This term is purely topological in $d=4$ dimension and doesn't have any dynamical effect but can offer interesting dynamics if it is non-minimally coupled with any other field such as a scalar field. In most inflationary models we need to consider a scalar field called inflaton which is responsible for the phenomenon of rapid expansion. One simplest way to generate inflation is to consider a minimally coupled scalar field with an unconventional equation of state. In the scenarios where the scalar field has a Higgs like potential, the observational constraints put stringent bound on these theories \cite{higgs}. For indirect constraints on cosmological
parameters from current Higgs mass value see \cite{riotto} . One of the main reasons to consider non-minimally coupled scalar field is it greatly modifies the spacetime dynamics, therefore, significantly improves the usual shortcomings of minimal
scenarios, such as usual chaotic and Higgs inflationary models. In this paper, we will consider specific class
of models where a scalar field couples with the Gauss-Bonnet term non-minimally. There have been a lot of studies on the implications of scalar coupled Gauss-Bonnet term in the context of inflationary scenario \cite{kawai, nepal, Koh:2014bka}\footnote{See also \cite{SBAC} for an alternate source of such interactions.}. In the context of dark energy model, Gauss-Bonnet term has also been considered \cite{sasaki2}.
Therefore, in the description of initial formalism and the background dynamics we will have significant overlaps of the current work with the 
aforementioned older ones. However as already mentioned, our main motivation is to put constraints from the current PLANCK data \cite{PLANCK}, and to try to understand the dynamics at a region where the Gauss-Bonnet inflaton coupling constant ($\beta$) is large.

With the increasing precision of cosmological experiments, it is important to check whether any proposed theory satisfies the bounds on the parameters that have been imposed from observation. In this study we will have two dimensional parameter space $(m,\beta)$, where $m$ is mass of the inflaton. For Higgs like potential, one has to replace $m$ by quartic coupling parameter $\lambda$. Here we have considered both chaotic and Higgs inflation with linear and quadratic inflaton coupling with Gauss-Bonnet term. Using the constraints from Planck, we 
have computed important cosmological parameters and show that even with the 
Gauss-Bonnet like non-minimal coupling,
usual Higgs inflation is ruled out by the latest PLANCK result for the scalar spectral index $n_s = 0.9682\pm 0.0062$, and the tensor to scalar ration $r < 0.07$ (for latest bound see \cite{latestr}).
On the other hand, Gauss-Bonnet coupling improves the prediction of usual chaotic inflation 
scenario for quadratic potential by reducing the value of tensor to scalar ratio $r$ significantly.
In this paper we have gone beyond this inflationary analysis 
and have tried to impose further constraints on the parameters considering other physical as well as theoretical analyses. We have studied constraints coming from the reheating predictions \cite{liddle}. These constraints get imposed in reheating scenario due to the evolution of observable scales starting from the inflation to the current epoch, and from the entropy conservation. Interestingly the analysis provides severe constraints on the parameter space. To see whether these models remain well defined as quantum effective theories at the energy scales considered and to check the consistency with the aforementioned constraints we have also analyzed the unitarity bound on the Gauss-Bonnet coupling parameter $\beta$. This study is done considering  the $2$ graviton $\rightarrow 2$ graviton scattering amplitude at tree level. This exclude some part of the parameter space which are otherwise cosmologically viable.

The paper is organized as follows. In section-II, we study the inflationary dynamics and computed cosmologically
relevant quantities $(n_s,r)$. We find the viable parameter space consistent with the recent cosmological observations.
We found that Higgs like potential even with the non-minimal coupling is strongly disfavoured. Therefore, in 
all the subsequent sections, we will only consider chaotic type models. 
In section-III, we extensively discuss the reheating constraints that is consistent with the evolution of cosmological scales, and
the reheating entropy density. Interestingly these indirect reheating constraints set the lower and upper limits of $(m,\beta)$
respectively. In section-IV, we compute the unitarity bound on Gauss-Bonnet-inflaton coupling parameter $\beta$ by
calculating the $hh\rightarrow hh$ scattering amplitude in flat space. Finally we conclude with some proposal for future works.

\begin{figure}[t!]
\includegraphics[width=5.00in,height=1.800in]{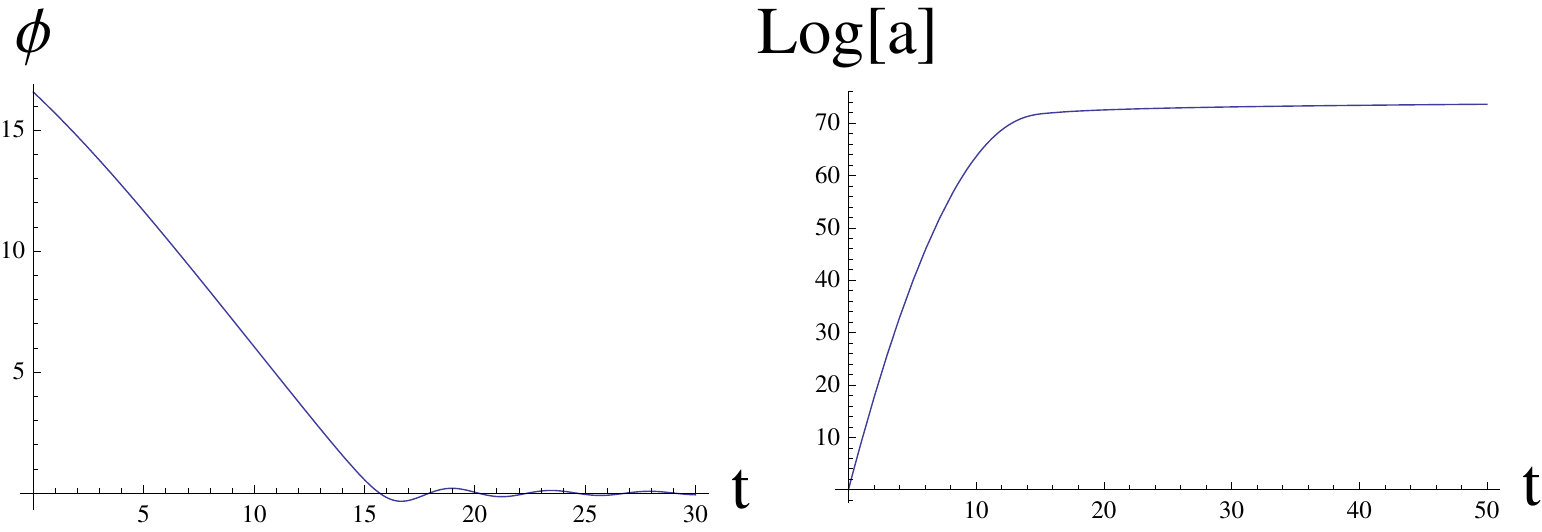}
\caption{\label{fig1} Typical inflationary background solution for Model-{\bf I}. We choose mass of the inflaton field
$m  \simeq 10^{-3} M_p$. Time is measured in unit of $m^{-1}$.}
\end{figure}

\section{The model: Background Evolution}

In this section we describe in detail the inflationary model with Gauss-Bonnet-scalar term.
We start with the following action, 
\bea
S = \int \sqrt{-g}[\frac {M_p^2} 2 R  - \frac 1 2
\pr_{\mu} \phi \pr^{\mu} \phi - V(\phi) - \frac {1} {16} F(\phi) L_{GB}]
\eea
where $L_{GB} = (R^2 -4 R_{\mu\nu}R^{\mu\nu} 
+ R_{\mu\nu\alpha\beta} R^{\mu\nu\alpha\beta})$, is the well know Gauss-Bonnet term.
$M_p^2= 1/(8 \pi G)$ is the reduced Planck mass. For our purpose we consider different possible forms of inflaton potential $V(\phi)$, and Gauss-Bonnet coupling function $F(\phi)$. 

The corresponding Einstein's equations of motion are
\bea
M_p^2 G_{\mu\nu} - \frac{\beta}{8}  P_{\mu \alpha \nu\beta} \nabla^{\alpha}\nabla^{\beta} F(\phi)  &=& T_{\mu\nu}\\
\Box \phi -  V'(\phi) + F'(\phi) L_{GB} &=& 0 
,\eea
where $H_{\mu\nu}$ is coming from $L_{GB}$, and $T_{\mu\nu}$ is the energy momentum tensor of the
inflaton field $\phi$. 
\bea
T_{\mu\nu} &=& \frac 1 2 \pr_{\mu}\phi \pr_{\nu}\phi - \frac 1 4 g_{\mu\nu} (\pr_{\mu}\phi
\pr^{\mu}\phi + V(\phi)) \nno\\
P_{\mu\alpha\nu\beta} &=& (2 R_{\mu\alpha\nu\beta} + 2 R_{\nu\alpha\mu\beta} -R g_{\alpha(}g_{\nu)\beta} + 2 R g_{\mu\nu}g_{\alpha\beta} -
4 R_{\mu\nu} g_{\alpha\beta} - 4 R_{\alpha\beta} g_{\mu\nu} + 4 R_{\alpha(\mu} R_{\nu)\beta}) \nno
.\eea
Where
$X = \frac 1 2 \pr_{\mu} \phi \pr^{\mu} \phi$ and $\Box = \frac 1
{\sqrt{-g}}\pr_{\mu}({\sqrt{-g}}\partial^{\mu})$.
 One of the interesting properties of the usual Gauss-Bonnet higher derivative term is that
it does not lead to any ghost degree of freedom. Therefore, even with the non-minimal coupling
with a scalar field, it is free of such spurious degree of freedom. 

Let us describe the study of the effect of the Gauss-Bonnet coupling term in the 
inflationary dynamics in light of recent cosmological experiment by PLANCK. We start with the spatially flat Freedman-Robertson-Walker(FRW) metric and the homogeneous inflaton background as
follows:

\be
ds^2 = -dt^2 + a(t)^2 (dx^2 +dy^2 + dz^2), \phi = \phi(t).
.\ee
Upon using the above Einstein's equations one gets the following dynamical equations for the scale factor $a$ 
\bea
3 M_p^2 H^2 =  \frac 1 2 \dot{\phi}^2  + V(\phi)  + \frac 3 2 H^3  F'(\phi) \dot{\phi},
\eea
and for the inflaton field 
\bea
\ddot{\phi} + 3 H \dot{\phi} + \frac {3} 2 H^2 (\dot{H} + H^2) F'(\phi) + v'(\phi) = 0
.\eea
Where, $H = {\dot{a}}/a$ is the Hubble constant.

In order to find the inflationary solution, we need to set the suitable initial condition. The strategy followed here is 
to identify the slow roll parameters which will set the correct initial condition out of infinitely many possibilities. The problem of initial condition in inflationary cosmology is well known and it has also been understood that it can not be answered in the framework of effective field theory. There exist significant effort to understand this issue from theoretical as well phenomenological point of view, For
a short but comprehensive review, see \cite{robert}. However, we do not address this issue here. As we have already mentioned, using the following slow roll parameters
\bea
\epsilon &=& \frac {M_p^2} 2 \left(\frac {V'}{V}\right)^2~~;~~\eta = M_p^2 \left(\frac
{V''}{V}\right) \nno \\
\alpha_1 &=& \frac {1} {4 M_p^2} {V' F'}~~;~~\alpha_2 = \frac {1} {6 M_p^2} {V F''} ~~~;~~~
\alpha_3 = \frac {1} {18 M_p^6} {V^2 F'^2},
\eea
the equations of motions for variables $(a(t), \phi(t))$ become 
\bea \label{aeq}
H^2 = \frac {V(\phi)}{3 M_p^2}~~~;~~~ \dot{\phi} = -\frac {1} 2 H^3 F'(\phi) -\frac {V'(\phi)}{3
H}
.\eea

Therefore, in order to have slow roll inflation one needs to make sure that all the slow roll parameters 
are less than unity till the end of inflation. The general procedure to obtain the inflationary solution is to derive 
the initial conditions for the inflaton field by using the following slow roll conditions,
\bea
\epsilon(\phi_f) &=& \frac {M_p^2} 2 \left(\frac {V'}{V}\right)^2 \Big{|}_{\phi_f} = 1~~~;~~~N(\phi)
= \int_{\phi_i}^{\phi_f}
\frac{6 H^2}{2 V'(\phi) + 3 H^4 F'(\phi)} d\phi = N_0,
\eea
where, $(\phi_i, \phi_f)$, are the values of inflaton field at the beginning and at the end of
inflation. We consider the value of e-folding number $N_0$ as one of the free
parameters. Depending upon the choice of its value, we will constrain our model parameters $(m,\beta)$. 
For the purpose our study we will consider various possible form of the potential and the
Gauss-Bonnet coupling functions as provided in the following table:    
\begin{center}
\[\begin{array}{|c|c|c|c|}
\hline
{\bf Inflation} & Type & V(\phi) & F(\phi) \\
\hline
&{\bf I} & m^2 \phi^2 & \frac{-\beta \phi}{M_p} \\
Chaotic & {\bf II} & m^2 \phi^2 & -\left(\frac{\beta \phi}{M_p}\right)^2  \\
\hline
& {\bf III} & \lambda \phi^4 & \frac{-\beta \phi}{M_p} \\
Higgs & {\bf IV} & \lambda \phi^4 & -\left(\frac{\beta \phi}{M_p}\right)^2 \\
\hline
\end{array}\]
\end{center}

For the sake of generality, we consider the Higgs potential as well. However, as we will
see, that current PLANCK data strongly disfavour this Higgs like potential.  
A typical background solution for the $(a(t), \phi(t))$ is shown in the figure \ref{fig1},
where we have chosen $N_0\sim 70$. Now, let us consider two categories of model as we have displayed in the above table,
and constrain the model parameter space based on the observed values of the scalar spectral 
index $(n_s)$ and the tensor to scalar ratio $(r)$.

\subsection{Perturbation: Constraints through $(n_s,r)$ }

Most important cosmological observables such as large scale structure, CMB are 
originated from the cosmological perturbations of quantum origin. 
During inflation those quantum fluctuations evolve in the inflationary background at all scales. 
All the structure that we see in our present observable universe are believed to 
be directly connected to those quantum fluctuations through coherent-de-coherent transition.
Those primordial fluctuations have been observed by PLANCK in the CMB as temperature 
fluctuations which is $\delta T \simeq 10^{-5}~~{^0}K$. Usual procedure to study those
quantum fluctuations is to understand the dynamics of purely metric fluctuations in unitary
gauge,
\bea
ds^2 = - N^2 dt^2 + (\delta_{ij} + h_{ij})e^{2 \psi}(dx^i + N^i dt)(dx^j + N^j dt) ~~~;~~~\delta \phi
=0,
\eea
where, $N$ is the usual lapse function, and $N^i$ is the shift vector. Those metric components
will provide the Hamilton and momentum constraints respectively. In the gauge specified above,
$(\psi, h_{ij})$ are the dynamical scalar and tensor degrees of freedom respectively.
Those are the dynamical degrees of freedom which contribute to the density perturbation and the gravitational wave
background in the subsequent cosmological evolution after the end of inflation. As we have already 
mentioned in the introduction, the detailed study have been done on the perturbation 
analysis of Gauss-Bonnet inflationary model. We, therefore, will quote the main results
which are of direct cosmological importance.

The quantities of direct cosmological importance are the two point correlation functions
of the above scalar, $(\psi)$, and tensor $(h_{ij})$ degrees of freedom in physically motivated
Bunch-Davis vacuum. The quantities which parametrize the above two correlation functions are
scalar spectral index $n_s$, and scalar to tensor ratio $r$. 
The expressions for all these aforementioned cosmological quantities are \cite{kawai}
\bea
 n_s \sim 1- 6 \epsilon + 2\eta + \frac {2\alpha_1} 3 + 2 \alpha_2 ~~~;~~~r \sim 16 \epsilon +\frac {32 \alpha_1}{3} + 4 \alpha_3
\eea

The current cosmological observations say that the value of $ N \gtrsim 50$. 
So this particular lower limit on ${\cal N}$ provides further constraints on the model parameters.
We have computed the above observable quantities for different types of models that we have considered before,

\begin{figure}
\centering
\subfloat[\label{fig2}]{%
  \includegraphics[width=0.40\linewidth, height=0.172\textheight]{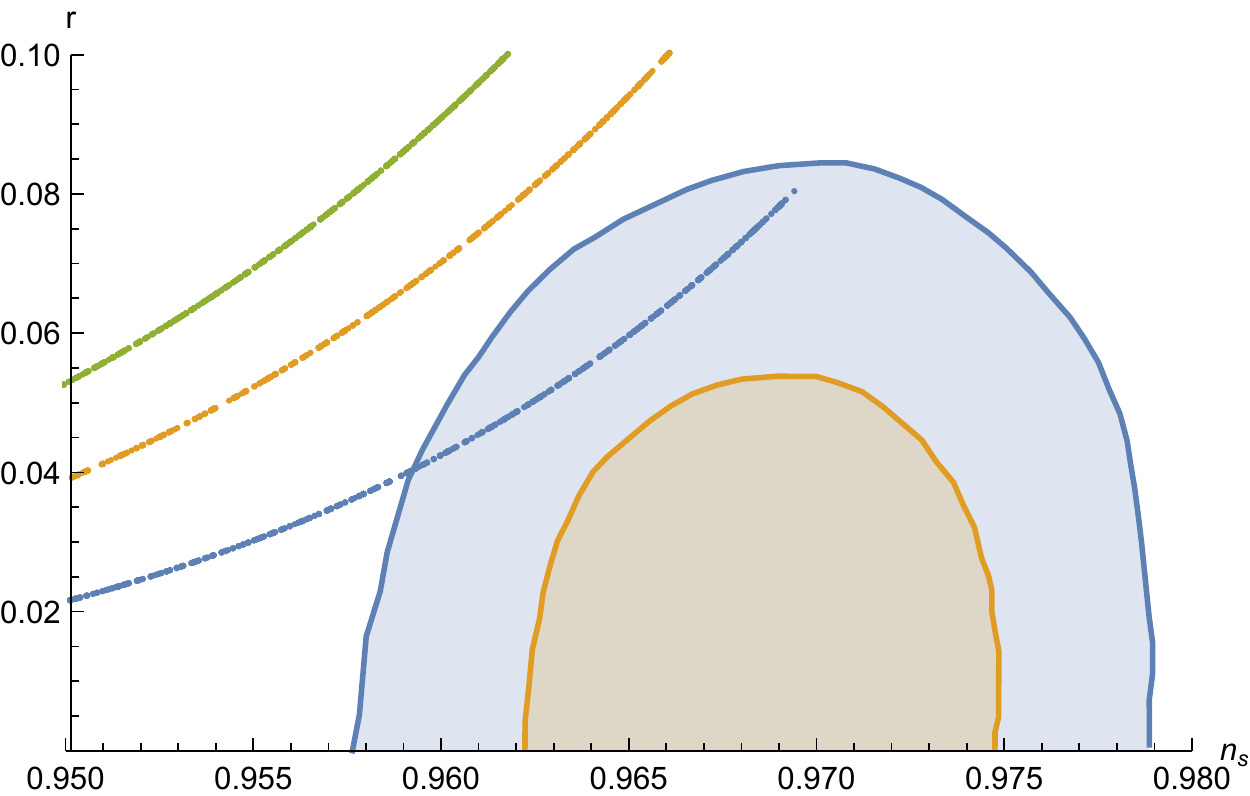}
}
\subfloat[\label{fig3}]{%
  \includegraphics[width=00.40\linewidth, height=0.172\textheight]{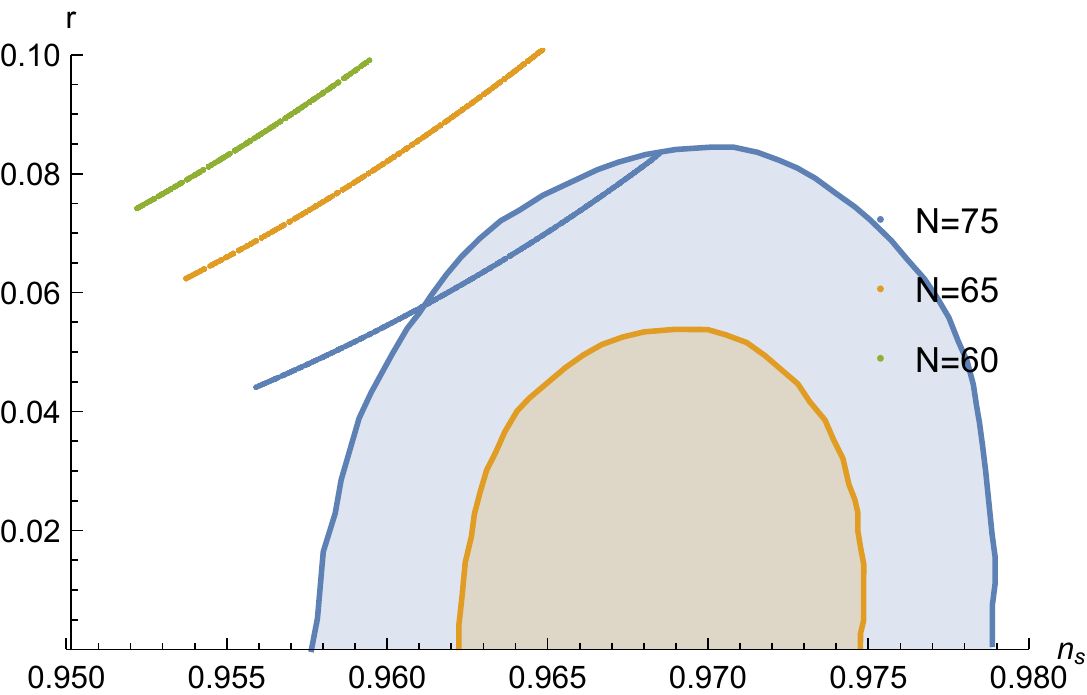}
}
\caption{$r~~ VS~~ n_s$ plot for chaotic inflation with a) Type-I and b) Type-II Gauss-Bonnet coupling
function. Inner and outer shaded regions are $1\sigma$ and $2\sigma$ constraints from Planck respectively. 
We have plotted three different e-folding numbers $N=60,65,75$ for green, orange and blue dotted lines respectively.}
\label{chaotic}
\end{figure}

In the fig-(\ref{chaotic}), we have considered the usual 
chaotic inflation (Type-I), where the inflaton field is coupled linearly with the Gauss-Bonnet term. We see that for a wide range of parameter values of $(\beta,m)$, the values of important cosmological parameters $(n_s,r)$ are well within
the limit of latest observation by PLANCK. It is apparent from these plots that increasing the number of e-folding makes the Type-(I,II) models
more consistent with the current cosmological observations. 
However, we have found a clear tension between constraints coming from the inflationary 
observables as discussed in this section and the constraints coming from the 
reheating prediction that we have discussed in the next section. 
Therefore, we find it very difficult to to figure out cosmologically viable parameter 
space which are consistent with both scenarios. 
Nevertheless, from the fig.(\ref{chaotic}a),
one can see that for Higgs type inflation, we could not find
viable parameter space. We get a wide range of parameter space in $(\beta,m)$ for both Type-(I,II) models as shown 
in fig.(\ref{chaotic}b). Within the cosmologically viable range, one finds a relation among $(m,\beta)$ of the form $\beta m^2 = q$,  
which is also evident from the expressions of the slow roll parameters. Here $q$ is a very small number in unit of Planck squared.
In our later discussion, we will see this particular combination of parameters will turn out to be the effective coupling
between the gravitational wave and the background inflaton field.
This fact will be very important for our subsequent study, specifically on unitarity which allows 
large $\beta$ only if the mass of the inflaton is small. 

\begin{figure}[t]
\centering
\subfloat[\label{fig2}]{%
  \includegraphics[width=0.40\linewidth, height=0.172\textheight]{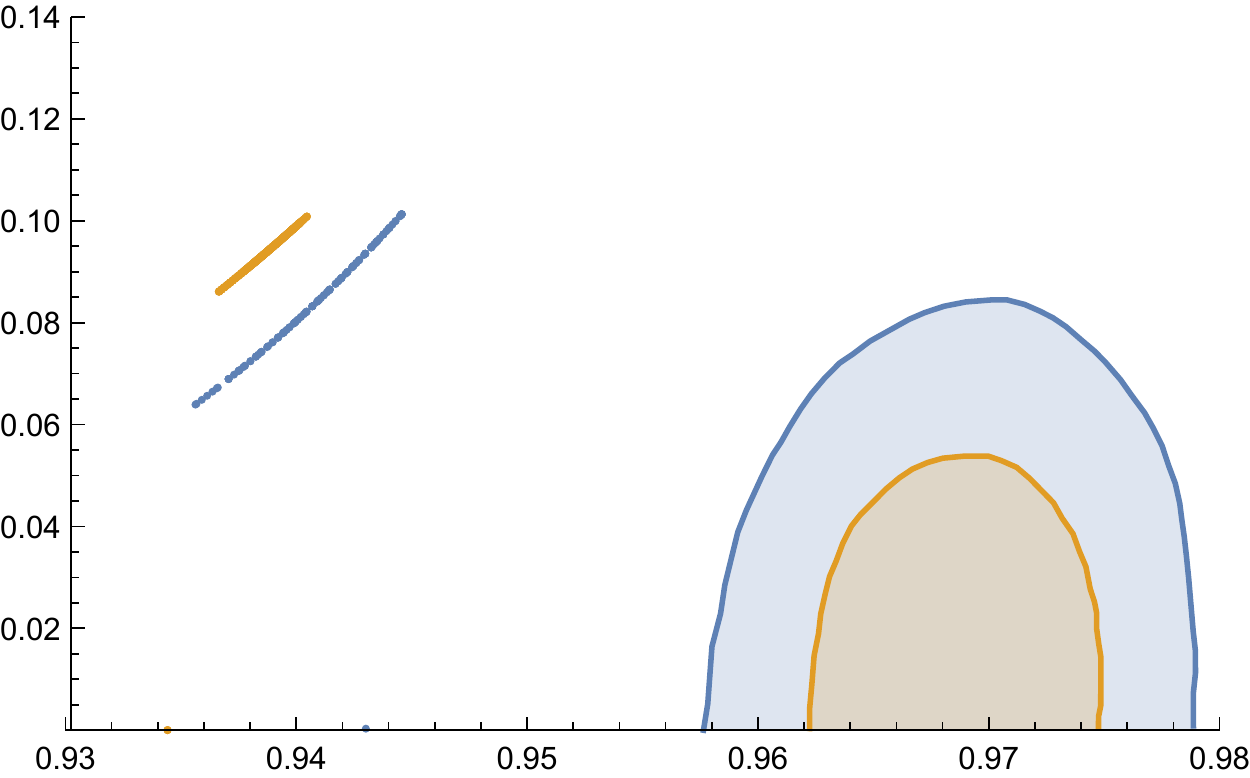}
}
\subfloat[\label{fig3}]{%
  \includegraphics[width=00.40\linewidth, height=0.172\textheight]{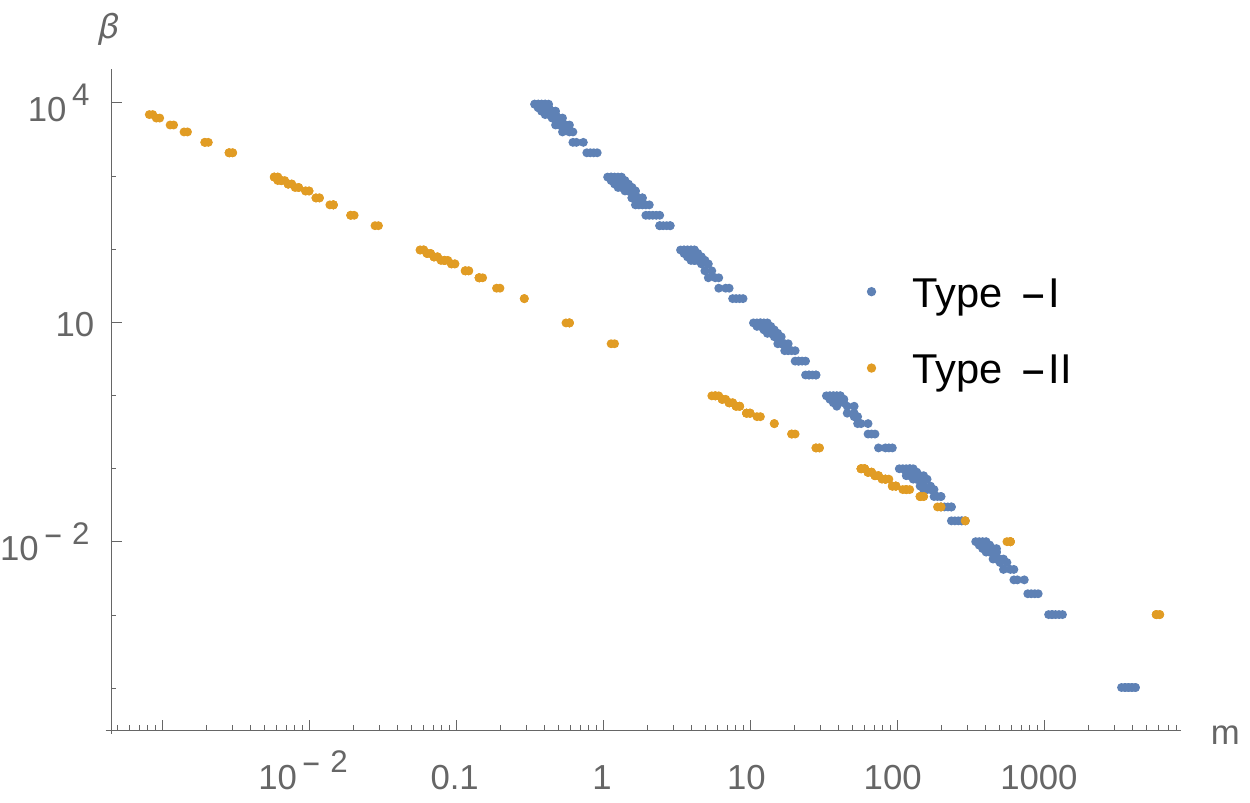}
}
\caption{a) Higgs inflationary model with linear inflaton coupling withe Gauss-Bonnet term. Therefore, it 
is with clear tension with the current Planck data. b) Allowed parameters in $(m,\beta)$ space for
chaotic inflationary models. $m$ is measured in $10^{-3} M_p$ unit.}
\label{higgs}
\end{figure}

\begin{figure}[t]
\centering
\includegraphics[width=0.80\linewidth, height=0.23\textheight]{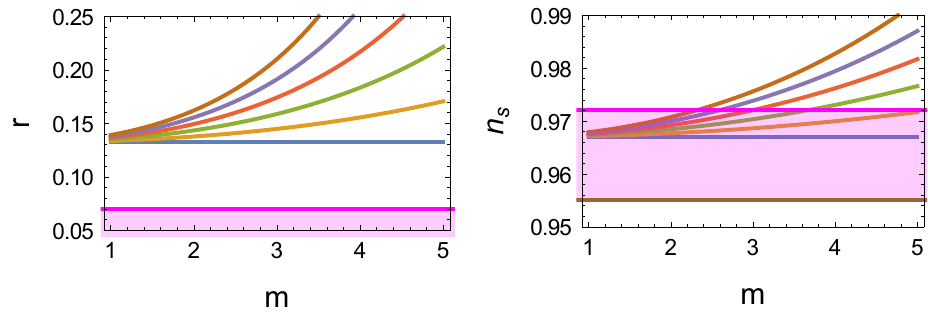}
\caption{a) Left panel: $(r ~vs~m)$ plot. b) Right panel: $(r ~vs~m)$ plot. For both the plot
we have considered only type-I model. We have plotted mass of the inflaton field in unit of 
$10^{-3} M_p$. Different curve corresponds to different value of negative $\beta =(0,-20,-40,-60,-80,-100)$. 
The horizon blue line corresponds to $\beta =0$, which is the usual chaotic inflationary case. Therefore,
as $\beta$ value increases in the negative direction, $(r,n_s)$ increases. Horizontal shaded regions are the 
experimental bounds.}
\label{nbeta}
\end{figure}

These analyses reveal, as long as the background dynamics and observed cosmological parameters $(n_s = 0.9682\pm 0.0062,r < 0.07)$ are concerned, 
chaotic inflation fig.(\ref{chaotic}) is favoured over the Higgs inflation as shown fig.(\ref{higgs}a). 
There exist a huge range of parameter space satisfying $\beta m^2 \simeq (10^{-4} , 10^{-8}) ~M_P^2$ for Type-(I,II) model
respective shown in fig.(\ref{higgs}b), chaotic inflation
predicts $r < 0.07$ which is not true for the pure chaotic inflation i.e. $\beta =0$ case. 
So far all we have discussed is for the positive $\beta$. At this point let us also comment on whether 
negative Gauss-Bonnet coupling is cosmologically allowed or not. In order to see, in the fig.(\ref{nbeta}), we have plotted the 
behaviour of cosmological observables $(n_s,r)$ with respect to $m$, taking different values of negative $\beta$. 
In the plot we have considered only type-I model, where the coupling function is linear in $\beta$.
For type-II model, the behaviour turned out to be the same. However, one can clearly
see that the observational constraint on $n_s$ allows some part of negative $\beta$ region. Interestingly those $\beta$ values
predict $r$ to be significantly above the observational limit $r<0.07$, fig.(\ref{nbeta}). 
It would be interesting to understand if there is any theoretical significance of this positivity of the scalar-Gauss-Bonnet coupling coming from the experimental observation.    

At this stage let us comment on the differences between the analyses that are present in existing literature on Gauss-Bonnet inflation \cite{kawai, nepal},\cite{Koh:2014bka} and in our paper. Two notable differences are the detail analysis of constraints coming from 
the consideration of the reheating phenomena and the unitarity analysis at the tree level
scattering amplitude specifically for large Gauss-Bonnet coupling parameter.
In \cite{Koh:2014bka} the authors have considered scalar-Gauss-Bonnet inflation 
with the same type of inflaton potential and the Gauss-Bonnet coupling. 
As long as the inflationary observables $(n_s,r)$ are concerned, 
our analysis so far does not have much difference except the fact that we have considered 
the latest cosmological observation from PLANCK.  
However, as we have emphasized earlier, our subsequent analyses offer major contribution to the existing literature. 
In the next section, we have analyzed the constraints coming from the evolution of scales of cosmological
importance and entropy density after the inflation.
Interestingly, we have found, it leads to severe restriction on the parameter space of the model.
We have also considered the tree-level scattering amplitudes of gravitons treating our theoretical model as an 
effective quantum field theory around flat background to find the upper bound on the coupling so that the 
theory doesn't loose unitarity. This study is important as this may constrain the parameter space further.
In unitarity analysis we have considered the large Gauss-Bonnet coupling with
the inflaton as this region of parameter space is most relevant here and has not be considered before.

\section{Constraining through Reheating predictions}
   Reheating period between the end of inflation and the beginning of the standard radiation 
phase is not well constrained by the cosmological observation. However, recently there has been 
an interesting development \cite{liddle,kamionkowski,cook} to distinguish various inflationary models by indirectly 
looking at the possible physically motivated reheating mechanism through the evaluation of a 
particular observable scale and the entropy density till today. In this section we will try to understand 
the possible physical mechanism which can be characterized by the number of e-foldings ($N_{re1}, N_{re2}$), 
Equilibrium temperature $T_{re}$ and the equation of state parameter $(\omega_{re1}, \omega_{re2})$ during the period of 
reheating. Here, we will do a small generalization of the previous method proposed in \cite{liddle}. We will
consider two step reheating process parametrized by two aforementioned equation of state parameters. 
Our approach will be little more realistic than the previous one. Although qualitative behaviour of those parameters 
$(N_{re}= N_{re1}+N_{re2}, T_{re})$ in terms of scalar spectral index will be same.   
Furthermore, we will see how the knowledge of the parameters $(N_{re1},N_{re2},T_{re},\omega_{re1},\omega_{re2})$ can lead us 
to understand the possible viable region of the parameter space of $(\beta, m)$. As we have seen, for the Higgs
inflationary model with Gauss-Bonnet coupling, we could not find viable parameter space which can produce successful inflation.
Therefore, in this section we will restrict our study on the the chaotic inflationary models. 
\begin{figure}
 \centering
        \includegraphics[width=00.7\linewidth, height=0.172\textheight]{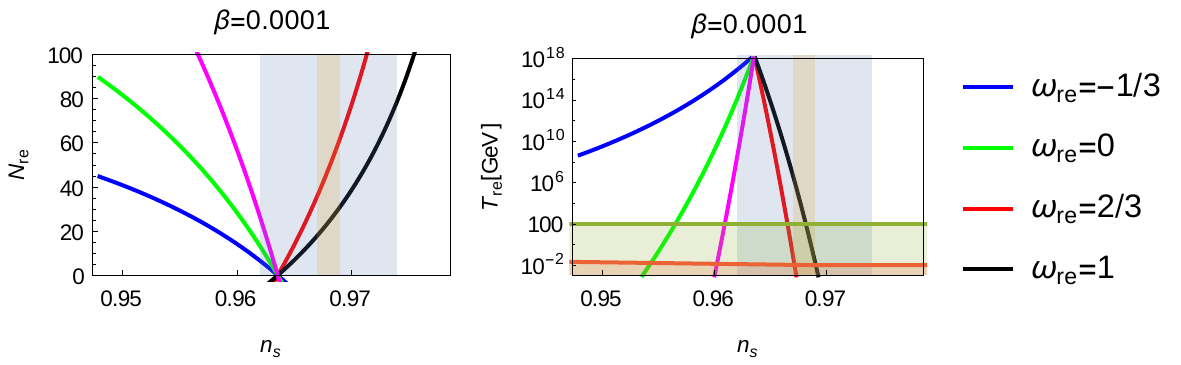}
        \caption{\scriptsize Variation of $(N_{re}, T_{re})$ as a function of $n_s$ have been plotted for 
$\beta = 0.0001$. Qualitative behavior of plot is same as those of no Gauss-Bonnet coupling.
 Green, blue, red, and black lines are the representative plots for 
the single reheating phase after the end of inflation. We consider four sample values of the equation of 
state parameters $\omega_{re} =(-1/3,0,2/3,1)$ during reheating. The magenta line corresponds to the two phase 
reheating process with the theoretically motivated set of equation of state parameters, 
$(\omega_{re1}=0,\omega_{re2} = 1/3)$, and equal number of e-folding parameters $N_{re1}=N_{re2}$.The light blue shaded region 
corresponds to the $1 \sigma$ bounds on $n_s$ from Planck. The brown shaded region corresponds to the $1 \sigma$
bounds of a further CMB experiment with sensitivity $\pm 10^{-3}$ \cite{limit1,limit2}, using the same central $n_s$ value as
Planck. Temperatures below the horizontal red line is ruled out by BBN. The deep green shaded region is below the 
electroweak scale, assumed 100 GeV for reference.} 
\label{beta0}
\end{figure}

Following \cite{liddle}, we find one of the important constraints comes from the relation between the scale $k$ that
we observe today and the same scale which exited the horizon during inflation. The equation which relates those scales,
$k = a_0 H_0 = a_k H_k$, provides us the following important relations among various e-folding numbers through out
the evolution of the universe,
\bea
\frac k {a_0 H_0} = \frac {a_k H_k}{a_0 H_0} = \frac{a_k}{a_{end}} \frac {a_{end}}{a_{re1}} \frac {a_{re1}}{a_{re}} \frac {a_{re}}{a_0}\frac {H_k}{H_0}
= e^{-N_k} e^{-N_{re1}}e^{-N_{re2}} \frac {a_{re} H_k}{a_0 H_0}.
\eea  
Another important relation among the reheating temperature $T_{re}$ and various e-folding numbers comes from the basic assumption that the reheating entropy is conserved through out the evolution from the radiation dominated phase to the current phase in the CMB and neutrino background. The conservation equation is as follows 
\bea \label{entropy}
g_{re} T_{re}^3 = \left(\frac {a_0}{a_{re}}\right)^3\left( 2 T_0^3 + 6 \frac 7 8 T_{\nu 0}^3\right).
\eea

\begin{figure}
\centering
\subfloat{
  \includegraphics[width=00.5\linewidth, height=0.172\textheight]{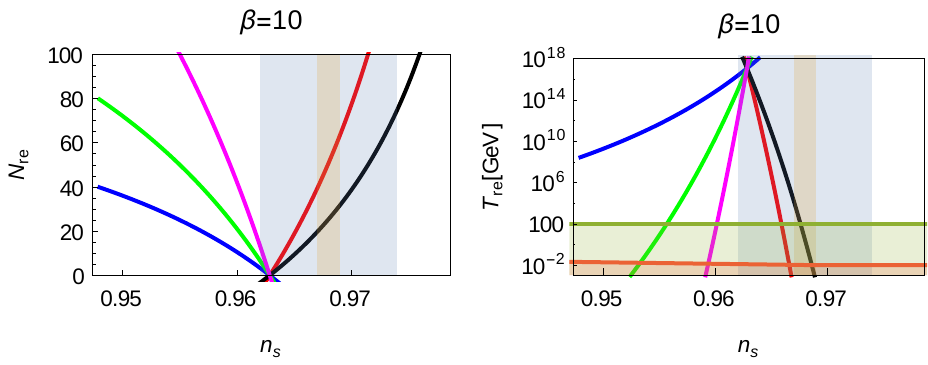}
}
\subfloat{
  \includegraphics[width=00.5\linewidth, height=0.172\textheight]{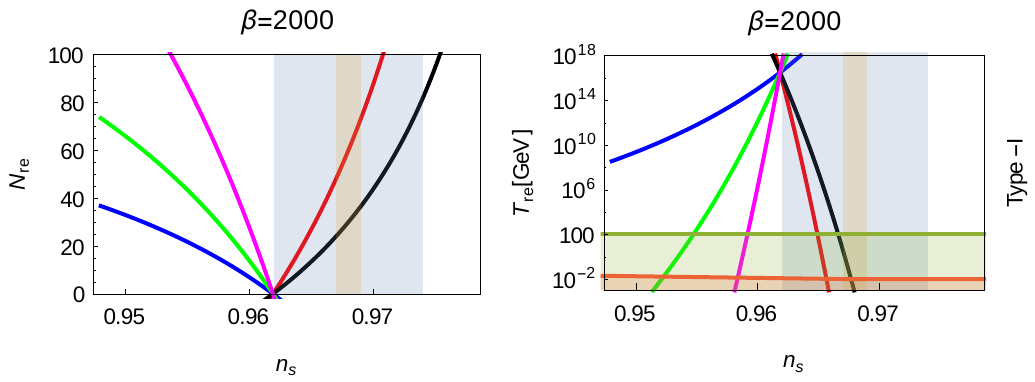}
}
\hspace{0mm}
\subfloat{
  \includegraphics[width=00.5\linewidth, height=0.172\textheight]{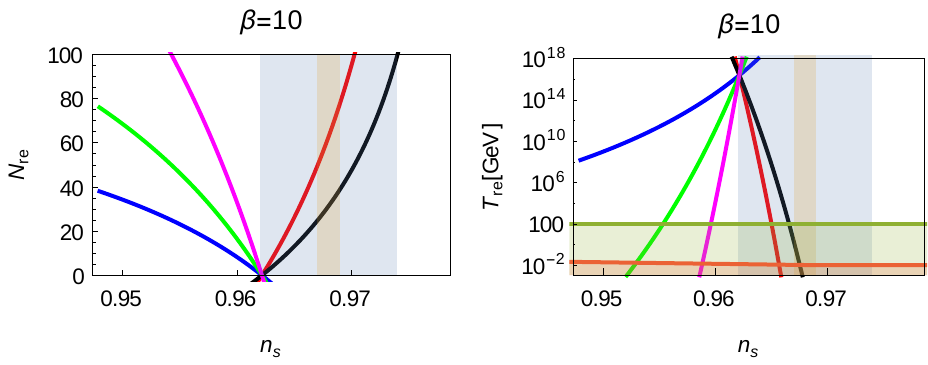}
}
\subfloat{
  \includegraphics[width=00.5\linewidth, height=0.172\textheight]{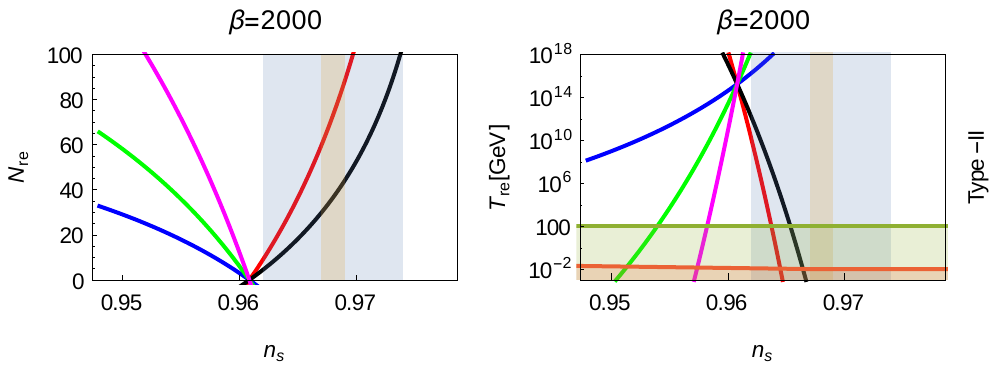}
}

\caption{\scriptsize Variation of $(N_{re}, T_{re})$ as a function of $n_s$ have been plotted for two different models 
with $\beta = (10,2000)$. All the other parameters are taken to be same as in the previous plot.} 
\label{beta10}
\end{figure}

Where, $(a_k, a_{end}, a_{re1}, a_{re}, a_0)$ are the corresponding values of the cosmological scale factor at the time of horizon exit of a particular scale $k$, 
at the end of the inflation, at the end of the first reheating phase with equation state parameter $\omega_{re1}$, 
at the end of the reheating phase with the equation state parameter $\omega_{re2}$, and at the present time respectively. $(N_k, N_{re1}, N_{re2})$ are 
the e-folding numbers parameterizing the relative expansion from $a_k$ to $a_{end}$, $a_{end}$ to $a_{re1}$, and during the second 
part of the reheating phase respectively. Therefore, according to the definition ${a_k}/{a_{end}} =e^{-N_k},  {a_{end}}/{a_{re1}}= e^{-N_{re1}}$ and 
${a_{re1}}/{a_{re}}= e^{-N_{re2}}$.
For the expression of conservation of entropy, $T_{re}, T_0, T_{\nu 0}$ are the reheating, present day's CMB and the neutrino 
temperature. We also know $T_{\nu 0} = (4/11)^{1/3} T_0$. $g_{re}$ is the effective number of degrees of freedom during 
reheating. From the above two master equation one can arrive at the following two expressions \cite{cook} among the important parameters
$(N_{re}, T_{re}, \omega_{re})$ which characterizes the re-heating phase after the end of 
inflation 
\begin{eqnarray} \label{nretre}
&& N_{re} = \frac{4(1+\gamma)}{(1-3\omega_{re1})+\gamma(1-3 \omega_{re2})}\left[61.6 - \ln\left(\frac{V_{end}^\frac{1}{4}}{H_k}\right) -N_k\right]\\
&& T_{re} = \left[\left(\frac{43}{11g_{re}}\right)^\frac{1}{3} \frac{a_0 T_0}{k} H_ke^{-N_k}\right]^{\frac{3(1+\omega_{re1})+\gamma(1+\omega_{re2})}
{(3\omega_{re1}-1)+\gamma(3 \omega_{re2}-1)}}
 \left[\frac{3^2.5V_{end}}{\pi^2g_{re}}\right]^{\frac{1+\gamma}{(1-3\omega_{re1})+\gamma(1-3 \omega_{re2})}}          .
\end{eqnarray}
During reheating
the scale factor will evolve according to the aforementioned equation of state parameters. 
In order to derive above relation, we parametrize, $N_{re2} = \gamma N_{re1}$.
Furthermore, we assume first phase of the reheating ends instantaneously. Therefore, during this phase one can relate various forms of 
energy densities by the following equation  
\bea
\frac {\rho_{end}}{\rho_{re}} = \left( \frac{a_{end}}{a_{re1}}\right)^{-3(1+ \omega_{re1})} \left( \frac{a_{re1}}{a_{re}}\right)^{-3(1+ \omega_{re2})}
\eea
All the above equations will turn into the same form which was discussed in \cite{cook}, if we consider $\gamma =0, \omega_{re2}=0$.
One can of course generalize the above analysis further by considering specific reheating model into consideration. We defer a detail
analysis of this for future study. As we can see from the above equation \ref{nretre}, all the quantity of our interest
can be calculated during the phase of inflation. As we have already solved the full background dynamics numerically in the previous
section, we use those numerical solution directly. We consider $k$ to be related to the pivot scale of PLANCK,
$k/a_0 = 0.05 Mpc^{-1}$, at which the scalar spectral index has been estimated to be $n_s = 0.9682 \pm 0.0062$.
Throughout our numerical calculation, we use the above pivot scale to constrain our model parameter.
For our numerical purpose, expression for all the important quantities in Eqs.(\ref{nretre}), are as follows
\begin{eqnarray}
H_k =  \sqrt{\frac{V(\phi_k)}{3 M_p^2}}~~~;~~~
N_k = \int_{\phi_{end}}^{\phi_{k}} \frac{6 H^2}{2 V'(\phi) + 3 H^4 \beta F'(\phi)} d\phi,
\end{eqnarray}
where, in the above expression, we use the slow roll approximation. $\phi_k$ and $\phi_{end}$ are the values of inflaton field 
at which a particular scale $k~ (\mbox{for our case}~ 0.05 Mpc^{-1})$ exits the horizon, and inflation ends respective.
We also calculated the value of $\phi_k$ from $n_s({\phi_k}) = n_s$ by
inverting it for different values of $n_s$. In Fig.(\ref{beta0}), we consider $\beta=0.0001$ for Type-I model, which predicts
almost the same values of $(N_{re}, T_{re})$  compared to the usual chaotic inflationary scenario \cite{liddle,cook} 
without Gauss-Bonnet coupling term. However, important point to note that for above mentioned value of $\beta$, 
$r$ is above the PLANCK bound specifically $r \simeq 0.07$ within the $1\sigma$ 
range of $n_s$, namely $0.9744 > n_s > 0.962$, for $m \simeq 2 \sim 4 M_p$. Therefore,
the energy scale of inflation would be very large. For this, unitarity should be checked as we have done in our later
section. However, it is worth emphasizing that the usual chaotic inflation ($\beta=0$), predicts 
$r>0.14$, which is anyway ruled out by PLANCK. Therefore, small but finite
value of $\beta$ improves the model by reducing the value of $r$.  In Fig.(\ref{beta10}), we have considered two chaotic models for $\beta =(10, 2000)$, and plotted the possible
values of $N_{re}$ and $T_{re}$ within the range of scalar spectral index $0.978> n_s >0.9475$.
For all the plots we considered four discreet set of values of the equation of state parameter $(\omega_{re1},\omega_{re2})=(-1/3,0,2/3,1)$ 
during reheating. However, it has been proved to be very difficult to construct an effective field theory
for $\omega_{re} >1/3$. Therefore, we try to put stronger bounds on all the parameters by 
considering realistic cases with reheating 
state parameter to be within $(0,1/3)$.
Other values of omega we kept for completeness. As those values may arise because of some exotic matter fields.
Through out the analysis we will try to place constraint on our model parameters by taking into account 
only $1\sigma$ region of $n_s$, namely $ 0.9744 > n_s > 0.962$, which is the vertical light blue shaded region.

One trend that one immediately notices, as we increase the value of $\beta$, all lines are shifting towards
lower $n_s$ value and going out of the $1\sigma$ region. Therefore, $\beta$ very high value is disfavoured by the PLANCK data.
Furthermore, for $\beta > 1$ , the Type-I model (linear inflaton-Gauss-Bonnet coupling) is more favoured than the Type-II model
(quadratic inflaton-Gauss-Bonnet coupling). In the table below, we have provided some sample values which
illustrate these points. However, within the cosmologically viable parameter space, and for simplest
two stage reheating phase with $(\omega_{re1} = 0, \omega_{re2} =1/3)$, and equal number of e-foldings for each phase $(\gamma =1)$,
the model naturally predicts very high reheating temperature $T_{re} > 10^{12}$, and even more importantly it 
favours the instant preheating scenario, namely $N_{re}$ is very small. Such a high reheating temperature could be 
interesting in the context of baryogenesis. 
One of our important motivations to re-analyze the inflaton-Gauss-Bonnet scenario is to understand
viability of the model in large $\beta$ regime, and to study the possibility of resonant gravitational wave production, that in turn may lead to gravity mediated reheating scenario based on an earlier work by one of the authors  \cite{debu}. However, we will see that this is not the case because of severe constraints 
coming from the slow roll condition. 

As one sees from Fig.{\ref{beta10}, specifically for Type-I model,
within a huge range of $\beta = (10, 2000)$, if we consider wide range of equation of states during reheating, 
predicted value of $(n_s, r)$ could be within the $1\sigma$ limit of the PLANCK observation. 

{\scriptsize 
\[\begin{array}{cc}
\text{Relevant parameters: First two rows are for Type-I, last row is for Type-II models}\\ 
\begin{array}{|c|c|c|c|c|c|}
\hline
\{\beta, m (10^{-3} M_p)\} & e-folding(N) & r & n_s & T_{re}(GeV) & N_{re}  \\
\hline
(0.002,~ 630) & 63.3 & 0.082 & 0.9612 & 6.1 \times 10^{17} & 1.2\\
  & 62.1 & 0.084 & 0.9607 & 2.6 \times 10^{14} & 10.0\\

(0.002,~ 700) & 63.2 & 0.075 & 0.9595 & 3.9 \times 10^{17} & 1.7 \\
  & 62.3 & 0.077 & 0.9591 & 6.6 \times 10^{14} & 14.9 \\
\hline
(20,~ 6.6) & 61.0 & 0.084 & 0.9595 &  1.2 \times 10^{17}  & 0.45 \\
 & 60.2 & 0.086 & 0.9591 & 3.4 \times 10^{14} & 7.1\\

(20,~ 7.0) & 60.94 & 0.08 & 0.958 &  6.5 \times 10^{16} & 1.2\\
 & 60.3 & 0.081 & 0.9582 &  7.5 \times 10^{14} & 6.3\\

\hline

(10,~ 0.45) & 59.63 & 0.097 & 0.9585 & 2.20 \times 10^{16} & 0.83\\

   & 58.27 & 0.100 & 0.9579 & 1.74 \times 10^{12} & 14.88\\

(10,~ 0.40) & 59.49 & 0.104 & 0.9601 & 1.32 \times 10^{16} & 1.52\\
     & 58.41 & 0.107 & 0.9596 & 6.41 \times 10^{12} & 10.06\\

(10,~ 0.35) & 59.51 & 0.110 & 0.9616 & 1.81 \times 10^{16} & 0.91\\
  & 58.28 & 0.113 & 0.9610 & 3.51 \times 10^{12} & 10.68\\
\hline
\end{array}
\end{array}
\label{table1}
\]}

In the table above, we have listed some sample values of the all the cosmologically viable
parameters and our model parameters for Type-(I, II) models. All those number are generated 
by considering the two-phase reheating scenario with $\gamma =1$. Form the above table we see that 
for Type-I model higher value of $\beta>1$ could be cosmologically viable within the $1\sigma$ range of $n_s$ from Planck, and it also predicts $r$ close to $0.07$ for $N\simeq 60$.
However, let us emphasize an important fact which is also evident from the above table, 
the reheating constraints severely restrict the possible values of $(n_s,r)$ so that the number of 
e-foldings during reheating $N_k$ should become positive. We also see this fact not only limits
the value of e-folding number during inflation closed to $N\simeq 60$ which is desirable but
it creates a clear tension between the set of values of $(n_s,r)$ with the experimental observations. However, as one can understand, 
this is clearly a model dependent conclusion in terms of extremely simplified assumptions of
complicated reheating process taking into account just background expansion and the equation of state. A more realistic and complete analysis may improve or rule out the Gauss-Bonnet modified inflationary scenario.
We will take up this issue in our future publication.     
From the last row of the above table it is apparent that the predictions of Type-II model are more constrained and in clear tension with the 
results of PLANCK within $1 \sigma$ region of $n_s$. In the next section, we will try to see if the unitarity constraint coming form 
$ hh \rightarrow hh$ scattering amplitude can support the higher value of $\beta$ during inflation specifically
for Type-I model.

\section{Analyzing Unitarity in scalar-GB theory}
So far we have discussed about constraining the scalar-Gauss-Bonnet coupling based on the classical cosmological 
background evolution as well as the quantum evolution of perturbation in the aforementioned background. For quantum theory,
energy scale of a certain physical process is important. In any inflationary cosmology it is the inflaton energy
density which sets the energy scale for the quantum evolution of the perturbation. 
As it happens in the effective field theory framework, any higher dimensional operator may lead to unitarity 
violation even below the natural cut off scale derived by simple dimensional analysis of a given interaction Hamiltonian. 
For example natural cut off scale is the Planck scale for the scalar-Gauss-Bonnet theory we are studying. 
Therefore, it is important to check if there exist any unitarity bound for a physical perturbative scattering process 
such as 2 graviton ~$\rightarrow$~ 2 graviton mediated by scalar field.
A perturbative process in quantum theory is valid only below the unitarity scale which can be
estimated by computing the scattering amplitude. In order to compute the amplitude, we can consider 
flat classical Minkowski background instead of inflationary background for computational simplicity.
However, validity of this approximation is usually explained by the fact that at the length scale
of scattering process the spacetime is approximately flat.  
In this section we will be calculating the tree level four point graviton scattering amplitude 
to set further constraints on $(m,\beta)$. For a comprehensive account on the issues related to the bounds imposed on the non-minimal couplings of different gravity models, see \cite{atkins}. It is known, to achieve adequate amount of density perturbations one usually needs to consider a large value of the coupling constant for models involving higher curvature couplings \cite{Salopek:1988qh, Fakir}. Here our intention is to determine the maximum limit of the coupling constant allowed by the unitarity consideration of the theory. Our 
analysis will be based on dimensional arguments used in field theory. 

We rewrite the action for a single scalar(inflaton) field coupled non-minimally with higher derivative Gauss-bonnet combination,

\be
\LL= \sqrt{-g}\left[\frac {M_p^2} 2 R  - \frac 1 2
\pr_{\mu} \phi \pr^{\mu} \phi - V(\phi) - \frac {1} {16} F(\phi) L_{GB}\right]
\label{Lag}
\ee

where $L_{GB}= R^2\,-4\, R^{\m\n}R_{\m\n}\,+\,R^{\m\n\a\b}R_{\m\n\a\b}$. In this section will consider
the Type-I chaotic scenario i.e $F(\phi) = \beta (\phi/M_P)$, where $\beta$ is dimensionless coupling constant. 
We analyze the tree-level $2$-graviton $\rightarrow 2$-graviton scattering amplitude.

We first expand the metric around flat Minkowski space upto second order in $\k=M_P^{-1}$.

\be
g_{\m\n}=\eta_{\m\n}\,+\,\k\, h_{\m\n}
\ee

Also,

\be
\sqrt{-g}= 1\,+\,{\k \over 2} h\,-\,{\k^2\over4}(h_{\m\n}^2-{1\over2}h^2)\,+\cdots
.\ee
We will adopt diag-$\eta_{ab}=(-1,1,1,1)$ signature. We now express the Lagrangian as a sum of free (kinetic) terms plus interaction terms as,

\be
\LL=\LL_{free}\,+\LL_{int},
\ee

using the gauge \be \partial_{\m} h^{\m\n}~=~{1\over 2}\partial^{\n} h. \label{gaugefixing}\ee

with \be\LL_{free}=-{1\over 4}\del_{\a}h_{\m\n}\del^{\a}h^{\m\n}\,+\,{1\over 8}\del_{\a}h\del^{\a}h \,+\, {1\over2}\del_{\m}\phi\del^{\m}\phi\,-{1\over2}m^2 \phi^2,\ee

where we have taken $V(\phi)$ to be a mass term only. We also write the expression for Riemann tensor after expanding around flat metric

\be
R^{\a}_{\b\m\n}={\k\over2}\left[\del_{\b}\del_{\m} h^{\a}_{\n} -\del^\a\del_{\m} h_{\b\n}-\del_{\b}\del_{\n} h^{\a}_{\m}+\del^{\a}\del_{\n}h_{\b\m} \right]
\ee
The interaction Lagrangian can be expressed as a series in $\k$ as:

\be
\LL_{int}=\LL_{int}^1\,+\,\LL_{int}^2\,+\,\cdots
\ee

with \[\LL_{int}^1=\fr{\k}{2}\left[-h^{\m\n}\del_{\m}\phi\del_{\n}\phi\,+\,{1\over2}h(\del_{\a}\phi\del^{\a}\phi\,-\,m^2\phi^2)\right]\]

and 

\begin{eqnarray}
\LL_{int}^2&=&\k^2\left[\zeta\phi\left({1\over4}(\Box h)^2-(\Box h_{\m\n})^2+(\del_{\m}\del_{\n}h_{\a\b})^2-\del_{\a}\del_{\m}h_{\b\n}\del^{\m}\del^{\n}h^{\a\b}\right)\right. \nonumber \\ &-& \left.{1\over4}(h_{\m\n}^2-{1\over2}h^2)({1\over2}\del_{\a}\phi\del^{\a}\phi-{1\over2}m^2 \phi^2)-{1\over2}hh^{\m\n}\del_{\m}\phi\del_{\n}\phi+\,\cdots\right]\label{int}
,\end{eqnarray}

where the dots will have terms from the expansion of $\sqrt{-g}R$ part of the action and we have set $\z=\b/16$. The relevant interaction for our study is the one with $\zeta$ parameter. This is a trivalent $\phi-h-h$ coupling and will give rise to the leading order diagram in $2$-graviton $\rightarrow 2$-graviton scattering amplitude or 4-point Green's function (Fig.$7$).

\begin{figure}
    \begin{center}
      \resizebox{30mm}{!}{\includegraphics{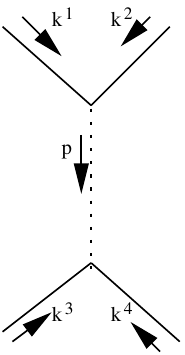}}
      \caption{Tree level diagram for $2$-graviton $\rightarrow 2$-graviton scattering in s-channel. The dotted line denotes scalar propagator. }
      \label{test}
    \end{center}
  \end{figure} 

The vertex for the corresponding interaction is sum of four terms given in the first line of (\ref{int}). Henceforth we call it collectively as $\LL_{\z}$. The Feynman rules for these four vertices are given by:

\be V^1_{\m\n \a\b}=i\zeta \k^2 \fr{\eta_{\m\n}\eta_{\a\b}}{4}k_1^2 k_2^2 ,\ee
\be V^2_{\m\n \a\b}=i\zeta \k^2 \fr{\eta_{\m\a}\eta_{\n\b}\,+\,\eta_{\m\b}\eta_{\n\a}}{2}k_1^2 k_2^2,\ee
\be V^3_{\m\n \a\b}=i\zeta \k^2 \fr{\eta_{\m\a}\eta_{\n\b}\,+\,\eta_{\m\b}\eta_{\n\a}}{2}(k_1.k_2)^2 ,\ee
\be V^4_{\m\n \a\b}=i\zeta \k^2\fr{k_{1\m}k_{2\a}\eta_{\n\b}\,+\,k_{1\n}k_{2\b}\eta_{\m\a}}{2}k_1.k_2.\ee

The scalar propagator is given by:

\be G(p)=\fr{1}{p^2\,+\,m^2}\ee

The external graviton legs will be associated with polarization tensors $\e_{\m\n}^{\l}$, with $\l=1,2$ and they satisfy

\be \e_{\m\n}^{\l}f^{\m\n,\a\b}\e_{\a\b}^{\l^{'}}=\delta^{\l\l^{'}}.\ee

$f_{\m\n \a\b}$ is the residue of the graviton propagator on-shell

\be f_{\m\n \a\b}=\eta_{\m\a}\eta_{\n\b}\,+\,\eta_{\m\b}\eta_{\n\a}-\eta_{\m\n}\eta_{\a\b}.\ee

and the graviton propagator is 
 
\be i D_{\m\n,\a\b}(k^2)=i\fr{\eta_{\m\a}\eta_{\n\b}\,+\,\eta_{\m\b}\eta_{\n\a}-\eta_{\m\n}\eta_{\a\b}}{2 k^2}.\ee

Since massless graviton can occur in two polarization states and it is a symmetric metric one can write the helicities in terms of product of two spin-one helicities. 

\be
 \e_{\m\n}^{1}=\e_{\m}^{+}\e_{\n}^{+}\,\,; \qquad \e_{\m\n}^{2}=\e_{\m}^{-}\e_{\n}^{-}
.\ee

Also it is easy to see due to the gauge constraint (\ref{gaugefixing}) these polarization vectors must satisfy the following relations,

\be \eta^{\m\n}\e_{\m}\e_{\n}=0\,\,;\qquad k^{\m}\e_{\m}=0.\ee

The scattering amplitude for two graviton to two graviton scattering with a scalar field at the internal line can be obtained from the following 4 point Green's function:

\be \zeta^2\k^4 \int d^4x d^4y <0|Th_{\m\n}(x_1)h_{\a\b}(x_2)h_{\r\s}(x_3)h_{\g\d}(x_4)\LL_{\z}(x)\LL_{\z}(y)|0>_c\ee

 After putting this expression to LSZ reduction formula we get the following expression for scattering amplitude in the particular gauge chosen: 

\begin{eqnarray} \M_{channel}=<k_3,k_4|k_1,k_2>=&-&i(2\pi)^4 \delta(k_1+k_2-k_3-k_4)\Sigma_{\l_1,\l_2,\l_3,\l_4=1}^2\fr{\z^2\k^4}{(k_1 + k_2)^2 - m^2}\nonumber\\ &\times&\left[-k_1^2k_2^2k_3^2k_4^2\e_{\m\n}^{\l_1}\e^{\m\n}_{\l_2}\e_{\a\b}^{\l_3}\e^{\a\b}_{\l_4}\right.\nonumber \\&+& \left. (k_1.k_2)^2(k_3.k_4)^2\e_{\m\n}^{\l_1}\e^{\m\n}_{\l_2}\e_{\a\b}^{\l_3}\e^{\a\b}_{\l_4}\right. \nonumber \\ &-&\left. 2(k_3.k_4)^2k_1^2k_2^2\e_{\m\n}^{\l_1}\e^{\m\n}_{\l_2}\e_{\a\b}^{\l_3}\e^{\a\b}_{\l_4}-2(k_1.k_2)^2k_3^2.k_4^2\e_{\m\n}^{\l_1}\e^{\m\n}_{\l_2}\e_{\a\b}^{\l_3}\e^{\a\b}_{\l_4}\right]\nonumber\\\end{eqnarray}

As on-shell gravitons are massless only the second term of the above expression survives. Thus the invariant scattering amplitude in the s-channel will have the following form, \footnote{The Mandelstam variables are expressed as:
\begin{eqnarray*} s&=& -(k_1 + k_2)^2=-(k_3 + k_4)^2 \nonumber\\ t&=&-(k_1 - k_3)^2=-(k_2 -k_4 )^2\nonumber \\ u&=&-(k_1 - k_4)^2=-(k_2 -k_3 )^2
\end{eqnarray*}}

\be \M_{s}=<k_3,k_4|k_1,k_2>=\Sigma_{\l_1,\l_2,\l_3,\l_4=1}^2\fr{\z^2\k^4s^4}{16(s - m^2)}\e_{\m\n}^{\l_1}\e^{\m\n}_{\l_2}\e_{\a\b}^{\l_3}\e^{\a\b}_{\l_4}
.\ee
 
The upshot of above analysis is the scattering amplitude scales as $\zeta^2 \fr{E^6}{M_P^6}$ in all channels if the mass of the inflaton field can be ignored with respect to the energy scale of the scattering process under consideration (note that $\phi$ is divided by $M_P$ to make $\beta$ dimensionless). This will enable us to determine the scale at which the unitarity of the theory will be under danger \cite{Trott, Hertzberg:2010dc}. In fact the theory has a cutoff below the Planck scale. This unitarity behaviour can be understood in the following way: For conformally coupled scalar theory like $R\phi^2$ and for $f(R)$ theories like $R^2$, we have seen from Hertzberg's analysis that the scattering amplitudes eventually become independent of the coupling constant when one sums up the contributions from all the channels \cite{Hertzberg:2010dc}. For the $R\phi^2$ theory the $2 \phi \rightarrow 2 \phi$ tree level scattering amplitude was studied setting the massless scalars on-shell. A similar analysis  for 
$R^2$ theory with $hh\rightarrow hh$ scattering yield same result. However, as we have seen above, for the theory that we are analyzing no such dramatic cancellation happens. This is probably because of the fact that both the theories $R\phi^2$ and $R^2$ can be rewritten as Einstein gravity plus a scalar field theory (in Jordon frame). Since Einstein theory is unitary and the scattering matrix element is invariant under field redefinitions one should expect that the same will be true for these cases as they are just alternative way of writing Einstein gravity with a minimally coupled scalar field. However, Gauss Bonnet gravity coupled to scalar Lagrangian cannot be recast in such a similar way. Therefore only in the limit $\zeta \rightarrow 0$, the theory should coincide with the Einstein GR and the dramatic cancellation of the tree level scattering process like in the case for $R^2$ theory will not happen here. We can see from the expression of the matrix element that the scattering amplitude scales as $\
\zeta^2 \fr{E^6}{M_P^6}$, assuming the scalar field mass $m << M_P$. This shows that the theory will violet unitarity at a scale $\Lambda \sim M_P/\z^{1/3}$ or at the Planck scale. This means we should not have too large value of $\z$ = $\b/16$ in order to have sufficient density perturbation in the inflationary phase.

Now we are in a position to constrain the value of $\beta$ from the analysis done in the earlier sections. In the scattering amplitude, energy $E$ of the external graviton is set by the inflationary energy scale. 
Assuming the energy scale of the scattering process $E$ to be the maximum energy scale of 
inflation, $E = V(\phi_{inf})^{1/4} = (m^2 \phi_{inf}^2)^{1/4}$, unitarity constraint entails
\bea
&& \fr{\b^2 E^6 }{16^2 M_P^6} \ll 1 \implies \b \ll  16\left(\frac {M_P}{E}\right)^3 = 16 \left(\frac {M_P^4}{m^2 \phi_{inf}^2}\right)^{\frac 3 4},\\ \nno
&& Or \\ \nno
&& \fr{\b^2 E^6 }{16^2 M_P^6} \ll 1 \implies (\b \tilde{m}^2)^2 \left(\frac {\tilde{\phi}_{inf}^3}{16^2\tilde{m}}\right)
 \simeq 10^{-8} \left(\frac {1}{16\tilde{m}}\right) \ll 1 
\label{unit}
\eea 
where, $\tilde{\phi}_{inf}, \tilde{m}$, are the initial value and the mass of the inflaton in unit of Planck. 
We have written down two different expressions for the constraint. From the first one, we can directly talk about constraint
on the value of $\beta$ for a given value of energy of the scattering process.
For example, considering a typical values of 
$\beta=10, \phi_{inf} = 15 M_P, m =7.5 \times 10^{-3} M_P$, such that all the cosmologically relevant parameters take 
$n_s =0.962,~ r=0.099,~ N = 60.4,~ T_{reh}= 1.3 \times 10^{15}~ \mbox{GeV}$, one finds the 
bound on $\beta < 4.2 \times 10^2$, which is indeed much greater than $\beta =10$. 
Generally speaking a value of $\beta \sim 10^4$ will render the theory non-unitary at an energy 
scale $\sim 10^{17}$ GeV which is well within the constraints set by COBE etc. The second expression is interesting due to the fact that, it is valid for all the parameter ranges
of $(m,\beta)$ which satisfies the Planck observation with $1\sigma$ range of $n_s$ and $r<0.07$ \cite{latestr}.
However at this point again we would like to emphasize that the above parameter space could be further 
restricted by using the reheating constraint discussed before. As we have studied before, 
for a huge region of parameter space it becomes inconsistent with the background evolution of cosmological 
scale and the entropy conservation. However, this conclusion is based on the simplified assumption 
on the reheating process after the inflation. We will take up this important issue for our future study.  
Nevertheless, to derive the above expression \ref{unit} we have used the fact that the value of $\tilde{\phi}_{inf}\simeq 15$ is almost independent
of $(m,\beta)$, and $\beta \tilde{m}^2 \simeq 10^{-4}$. It is interesting to see the 
emergence of a new scale $\sqrt{\beta m^2} \simeq  10^{-2} M_p$, same as GUT scale,
which controls the unitarity bound for the Type-I chaotic model. 
Therefore, unitarity constrains the value of $\tilde{m} > 10^{-7}$, while keeping $\beta \tilde{m}^2$ fixed, which implies $\beta < 10^{10}$. However, as we have seen before, reheating constraints excludes most of the parameters space and limits $\beta$ even less than $10^3$ (see fig.(\ref{beta10})).

We are not analyzing the unitarity for the Type-II case namely $\xi(\phi)=\phi^2$. In fact, no $2$ graviton $\rightarrow 2$ graviton scattering happens at the tree level for this theory (it happens at one-loop level) and thus we cannot say anything about the unitarity from the similar analysis done above. However, power counting estimates of scattering processes suggest that high energy behaviour of this theory will be sickened. This may indicate scale at which unitarity of Type-II model will be lost is earlier than that of Type-I model. A concrete analysis of unitarity for this model will reveal the exact fact which we postpone for future.

\section{Gravitational wave dynamics during reheating}
In this section we discuss about the dynamics of gravitational wave during reheating especially for large $\beta$. It is known that after the end of inflation, the inflaton will have coherent oscillation leading to the
reheating phase of our universe. We will be following the discussion of \cite{debu}, and try to understand if there is a possibility of resonant production of gravitational wave which in turn will trigger a gravity mediated preheating. Our initial hope was this will indeed occur in this model also. However, as we will see in the following discussion the inflationary slow roll conditions severely constrain the parameter space in such a way that the effective Gauss-Bonnet coupling of the background inflaton field with the gravitational wave is suppressed by an amount $\beta m^2/M_P^2 \simeq 10^{-4}$. Choosing the following transverse and traceless gauge condition for the tensor perturbation,
\bea
\partial_i h^{ij} = 0~~;~~\delta_{ij} h^{ij} = 0, 
\eea
and the Fourier mode of tensor fluctuation $h_{ij}$,
\bea \label{gwmode}
h_{ij}(t,x) = \int \frac {d^3 {\bf k}}{(2 \pi)^{3/2}} \sum_{s=1,2} [e^s_{ij}({\bf k}) {\tilde h}^s ({\bf k},t) e^{ i {\bf k}\cdot{\bf x}} + h.c.],
\eea
the gravitational mode satisfies the following linear equation
\bea \label{gwave}
\ddot{{\tilde h}}^s + \left(3 H  + \frac {\dot{\cal U}} {\cal U}\right)\dot{{\tilde h}}^s + \frac {k^2}{a^2 {\cal U}}\left(1 - \frac{1}{M_P^2} \ddot{F}(\phi) \right)
 {\tilde h}^s =0 .
\label{grav}
\eea
where, $e^s_{ij}({\bf k})$ are the two independent polarization tensor of the gravitational wave.
Where, ${\cal U} = 1 - H\dot{F}(\phi)/(2 M_P^2 a^2)$. This is a modified Mathieu equation with the oscillating inflaton background. Now let us see if this leads to a resonant graviton production in certain
band of frequencies related to the frequency of the oscillation. We write the above
eq.(\ref{grav}) in terms of an appropriate time variable taken in the unit of $m^{-1}$, which is the natural oscillation time scale of the inflaton,
\bea \label{gwave}
\ddot{{\tilde h}}^s + \left(3 H  + \frac {\dot{\cal U}} {\cal U}\right)\dot{{\tilde h}}^s + \frac {k^2}{ m^2 a^2 {\cal U}}
\left(1 - \frac{\beta m^2}{M_P^2} \frac{\ddot{\phi}}{M_P} \right)
 {\tilde h}^s =0 .
\label{ubuy}
\eea

\begin{figure}[t!]
\includegraphics[width=5.00in,height=1.800in]{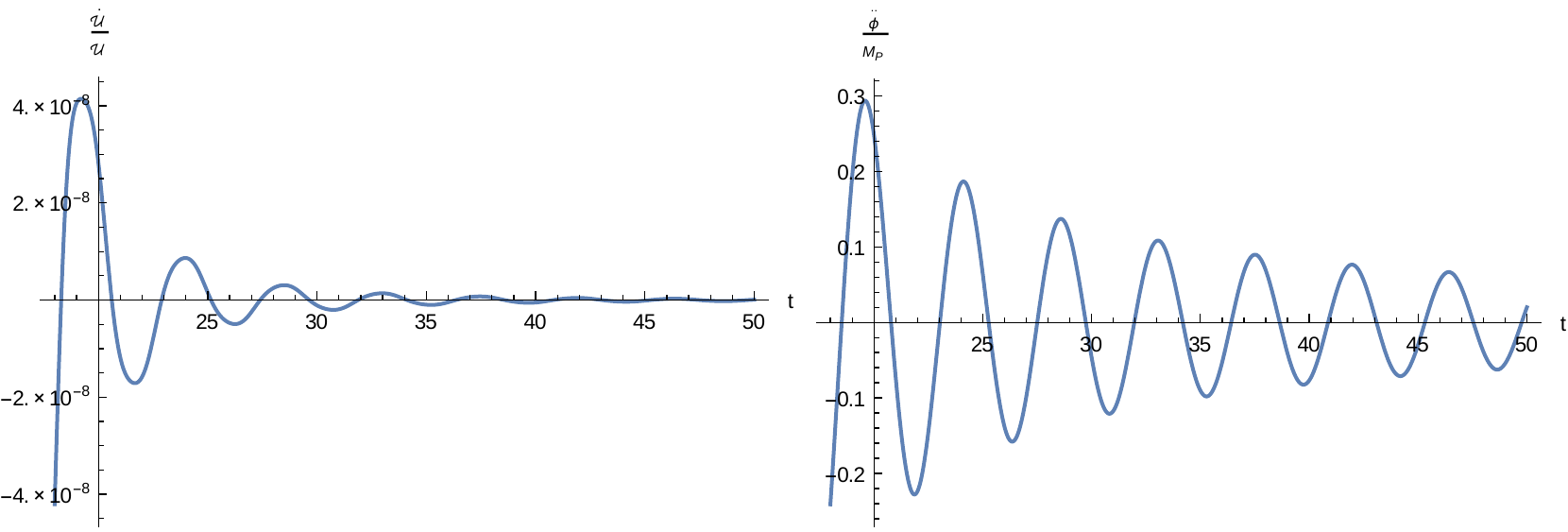}
\caption{\label{fig1} The oscillatory inflaton dependent effective mass term (right panel) and damping term (left panel) in the dynamics of the gravitational field during the reheating period. Time is measured in units of $m^{-1}$.}
\end{figure}

Clearly we can see that the non-minimal inflaton-Gauss-Bonnet coupling provides the oscillatory inflaton dependent
mass term and damping or anti-damping term in the dynamics of graviton during reheating period, as shown in Fig. (\ref{ubuy}). One notices that the amplitude of the dimensionless oscillatory effective mass term, and the damping terms are very small. In addition, as we emphasized before, slow roll condition sets the effective coupling parameter $\beta m^2/M_P^2 \simeq 10^{-4}$. Therefore, our numerical solution does not show any parametric resonant phenomena in the gravitational wave dynamics. Hence gravity mediated preheating will not be effective in this present case. This shows one has to resort to the usual reheating mechanism \cite{kofman}. However, another way of realization of reheating may occur here. We have seen from our earlier analyses on constraints arising from the reheating phase that the value of the reheating temperature increases with the increasing value of $\beta$. In fact the reheating temperature may rise as high as $T_{re}\sim 10^{10}-10^{16}$ GeV. This indicates instant pre-heating 
could 
also be important in order to understand the high re-heating temperature.

\section{Conclusions}
 
In this paper we have revisited the inflaton-Gauss-Bonnet model and studied in detail the cosmological as well theoretical aspects of it. The main goal of our work is to constrain the model parameters coming from the more recent cosmological observation made by PLANCK. 
We considered chaotic and Higgs type inflationary scenarios with two different kind of Gauss-Bonnet coupling. 
Considering the current cosmological observation by PLANCK, we found that Higgs type potential with non-minimal Gauss-Bonnet
coupling is not favoured at all.  At this point we would like to remind the reader that,
we have not considered usual Higgs inflation scenario, where $\zeta \phi^2 R$ \cite{higgs} type coupling is
introduced to effectively reduce the gravitational coupling which usually produces large
primordial scalar fluctuations. In our case we have considered Higgs-Gauss-Bonnet coupling, and it does not
provide sufficient amount of primordial fluctuations in accord with the cosmological observation.
However, the chaotic inflationary scenario can be significantly improved by the 
introduction of non-zero $\beta$. In addition to the production of sufficient scalar fluctuations, it also suppresses 
the tensor mode fluctuations which can be compatible with the experimental bound. It is important to remind the reader 
that without $\beta$ chaotic inflation is generically disfavoured because of its large prediction of tensor fluctuations. 
Therefore, introduction of non-zero $\beta$ not only improves the model from the cosmological point of view but also 
opens up new rich structures from the theoretical point of view. One such important aspect is its high energy
behaviour. In this paper we have particularly emphasized on the large $\beta$ region. One of the important reasons behind 
studying the large $\beta$ region is that the energy scale of inflation becomes few order of magnitude lower than 
the Planck scale in the cosmologically viable parameter ranges. Therefore, the effective field theory description will become more trustable.
Our analyses show that in order to satisfy observational constraint, one has to satisfy
$\sqrt{\beta \tilde{m}^2} \simeq (10^{-2}, 10^{-4})$ for Type-(I,II) chaotic models respectively.   
In order to constrain further, we studied the constraints coming from reheating and unitarity.  
We incorporated the reheating constraints coming
from the consistent evolution of cosmological scales and the conservation of entropy density.
We have invoked two step reheating process to obtain the relation between reheating 
temperature $(T_{re})$, the number of e-folding during reheating $(N_{re})$, and
various equation of states $(\omega_{re1},\omega_{re2})$. We leave the study of relations among those reheating parameters 
with a specific reheating models for future study. 

After obtaining the constraints from reheating analysis, we get severe restrictions on the possible values of $(n_s,r)$.
This is coming mainly from the positivity of the e-folding number $N_k$ during reheating.
This in turn severely restricts the parameter space of $(m,\beta)$.
Although the inflationary analysis provides a large viable parameter range in the from of 
$\beta m^2 \simeq (10^{-4}, 10^{-8})$, our analysis of reheating constraints makes it very difficult for type-II model to be viable. For type-I model also, it reduces the parameter space into the region
where $\beta$ is very small but mass of the inflaton $m$ is large. However, it is important to remember that the governing eqs.\ref{nretre}
is not completely model independent. In fact several important assumptions were made to parametrize the complicated 
reheating phenomena by few parameters. A more realistic and complete analysis may make the model viable or rule out the Gauss-Bonnet modified inflationary scenario.
We will take up this issue in our future publication.     

In this note we have calculated unitarity constraints by computing the tree level $hh\rightarrow hh$ scattering amplitude. For type-I model, we have found an interesting condition, $10^{-8} \left(\frac {1}{16\tilde{m}}\right) <<1$,
which in turn constrains the value of $\beta < 10^{10}$, providing the fact that at every value
of $\beta$, one satisfies the slow roll condition of inflation. 

One of the motivation to consider higher value of $\beta$ is to explore the 
possibility of resonant gravitational wave production during reheating. However, the model under consideration fail
 to produce sufficient amount of parametric resonance in order to trigger gravity mediated preheating. This happens because of the the parameters of these theories get constrained and have become not suitable for such scenario. This also shows the importance of examining any cosmological model with respect to all dynamical phases starting from inflation, both from the theoretical and observational point of view. However, there may exist other ways to realize (p)reheating with the scalar coupled Gauss-Bonnet terms. As we have already mentioned, natural preheating or instant preheating scenarios may be explored to see if they can produce the desired outcome. We plan to explore these scenarios in future. We are also planning to study other higher curvature theories that may be accommodated in the inflation-reheating scenario compatible with the present observational data.

\subsection*{Acknowledgments}
DM thanks the HEP-group members of IIT Guwahati to have vibrant academic discussions. SB thanks Shamik Banerjee for an illuminating discussion. S. B. acknowledges the support of Indian Institute ofInformation Technology, Allahabad where part of this draft was completed.

\end{document}